\documentclass[prb,aps,reprint,twocolumn,showpacs,
tightenlines,amsmath,amssymb]{revtex4-1}
\usepackage{graphicx}    
\usepackage{amsmath}     
\usepackage{amssymb}  
\usepackage{mathrsfs}
\usepackage{bm}                                          
\usepackage{colordvi}         
\usepackage{epsfig}
\newcommand{\bgreek}[1]{\mbox{\boldmath$#1$\unboldmath}}
\begin{document}

\title{
Spin-charge separation in bipolar
spin transport in
(111) GaAs quantum wells}
\author{M. Q. Weng} \email{weng@ustc.edu.cn}
\author{M. W. Wu} \email{mwwu@ustc.edu.cn}

\affiliation{Hefei National Laboratory for Physical Sciences at Microscale and
  Department of Physics, University of Science and Technology of China, Hefei,
  Anhui, 230026, China} 
 \date{\today}

 \begin{abstract}
   We present a microscopic theory for transport of the spin polarized charge
   density wave with both electrons and holes in the $(111)$ GaAs
   quantum wells. 
   We analytically show that, contradicting to the commonly accepted
   belief, the spin and charge motions are bound together only in
   the fully polarized system but can be separated in the case of
   low spin polarization or short spin lifetime even when the
   spatial profiles of spin density wave and charge density wave
   overlap with each other. 
   We further show that,
   the Coulomb drag between electrons and holes
   can markedly enhance the hole spin diffusion if the hole spin
   motion can be separated from the charge motion. In the high spin
   polarized system, the Coulomb drag can boost the hole spin
   diffusion coefficient by more than one order of magnitude.
 \end{abstract}


\maketitle

\section{Introduction}
The spin dynamics and spin 
transport in semiconductor nano-structures have attracted much
attention in the past decade 
due to the potential
applications of the spintronic devices.\cite{awschalomlossSamarth,
zuticrmp,*Fabianbook,dyakonovbook_08,Wu201061}
The spin transport in semiconductors can be described by a set of key 
parameters such as the spin relaxation time
$\tau_s$, spin mobility $\mu_s$ and spin diffusion coefficient (SDC)
$D_s$. These parameters can be directly extracted from the temporal
evolution of optically excited spin density wave
(SDW).\cite{PhysRevLett.76.4793,
*nature.437.1330,*weber:076604,
PhysRevB.82.155324,*PhysRevLett.106.247401,*Yang2012,
weng:063714,*PhysRevB.86.205307}
Since the optically excited electrons are
inevitably accompanied by an equal number of 
positively charged holes, the interplay between charge diffusion and
spin diffusion of electrons and holes is an important problem.

So far, the spin and charge dynamics are usually studied separately
by using different experimental setups. 
In the transient grating experiments, one can study the ambiplor
transport by creating a pure charge density wave (CDW)\footnote{It
  should be noted that the terms charge density wave, spin density
  wave and spin polarized charge density wave used in this paper are
  not the collective states of the electron system but the optically
  excited charges and/or spins whose densities change periodically in
  space.} 
without spin polarization using two parallel linearly polarized laser
beams.\cite{PhysRevLett.76.4793,*nature.437.1330,
*weber:076604,PhysRevB.82.155324,*PhysRevLett.106.247401,*Yang2012}
In the transient spin grating experiments, however, one studies the
spin dynamics by using two orthogonal linearly polarized laser
beams to create the mixture of a pure electron and hole SDWs with
spatial homogeneous carrier 
concentrations. The problem is further simplified in 
$(001)$ or $(110)$ GaAs quantum wells (QWs), since the hole spins
quickly disappear in subpicoseconds, leaving only the electron SDW 
together with the spatial homogeneous and unpolarized hole
gases.\cite{awschalomlossSamarth,zuticrmp,*Fabianbook,dyakonovbook_08,Wu201061}
The amplitude decay rates of the electron SDW is characterized by electron
spin lifetime $\tau_s^e$ and SDC $D_s^e$. 
In $(111)$ GaAs QWs, the electron and hole SDWs can coexist over a
few hundred picoseconds, as the hole spin lifetime can be comparable or
even larger than that of electrons in the special case of 
the cancellation of the Dresselhaus
and Rashba terms in the spin-orbit coupling
(SOC).\cite{PhysRevB.71.045313,vurgaftman:053707,sun:093709,PhysRevLett.107.136604,PhysRevB.85.235308}
It is expected that $(111)$ GaAs QWs would provide unique
playgrounds to show the rich physics when the spin and charge
degrees of freedom of both electrons and holes interplay with each
other comparing to $(001)$ or $(110)$ QWs. 

Recently, there are some experiments on the dynamics and transport in
the case when the charge and spin inhomogeneity coexist. 
In the transmission-grating-photomasked transient spin grating
experiments, a spin polarized carrier grating is generated by a
circularly polarized laser beam.\cite{oe.20.008192,*oe.20.003580}
Similarly, a spin polarized charge
package can be generated by a tightly focused circularly polarized
laser beam.\cite{PhysRevB.79.115321}
In the system where the spatial profiles of 
the SDW and CDW overlap, the spin and charge diffusions of
both carriers are assumed to be governed by the ambipolar diffusion,
since the spins are attached to the
carriers.\cite{oe.20.008192,oe.20.003580,PhysRevB.79.115321}
The SDC
can be further reduced and become smaller
than the charge diffusion coefficient (CDC)
if the spin Coulomb drag is taken into
consideration.\cite{amico_00,*badalyan_08,nature.437.1330,takahashi_08,jiang:113702,weng:063714} 
However, the validity of the assumption that spin and charge motions
are bound together in the case of spatial overlapping of SDW and CDW 
has never been justified beyond hand-waving
arguments.\cite{oe.20.008192,oe.20.003580,PhysRevB.79.115321}
When the carriers are fully spin polarized, the motion of the spins are
identical to that of the charges. In this case, it is expected that
the motions of spins and charges are indeed bound together. However,
it is not necessary so when carriers have different spins. 
A simple example is that when there are two electrons with opposite spins
moving with the same velocity but in opposite directions, there is 
a spin current but no overall charge current. Therefore, the spin
motion can be separated from the charge motion when the system is
not fully polarized. 
In the charge homogeneous system where pure spin
current can flow without the accompanying of a charge current,
the spin-charge separation (SCS) is almost taken for granted. 

For the spatially overlapped CDW and SDW, the
interplay between the charge and spin diffusion is more complex. 
Unfortunately, a solid theoretical investigation on the effect of
the interplay on the spin and charge motions from microscopic
approach has not yet been carried out up to date. 
The possibility of the SCS
in such system has not yet been explored.
In this Letter, we study the transport of a spin polarized CDW (SPCDW),
which can be excited optically\cite{oe.20.008192,oe.20.003580} and
has both charge and spin spatial inhomogeneities, in GaAs
(111) QWs, where both electron and hole spins can survive long
enough\cite{PhysRevB.71.045313,vurgaftman:053707,sun:093709,PhysRevLett.107.136604,PhysRevB.85.235308} 
to show the effect of the interplay between the charge and spin. 
We show that, contradicting to the currently accepted
notion,
the spin and charge motions are bound together only in the fully
polarized system, but can be separated in the system with low spin
polarization or short spin lifetime.
We further demonstrate that, in the case of the SCS 
the Coulomb drag between electrons and holes 
greatly enhances the hole spin diffusion 
since the electron spins move much faster than the hole spins.
When the spin polarization is high enough, the spin Coulomb drag
can boost the hole SDC
by more than one order of magnitude. 

\section{Model} 
We study the transport of spin polarized electron and  hole gases in
nonmagnetic GaAs QWs grown along the $z$-axis. One can write down the 
kinetic spin Bloch equations (\mbox{KSBEs}) 
as following,\cite{Wu201061,PhysRevB.66.235109,*PhysRevB.69.245320}
\begin{eqnarray}
  \label{eq:KSBE}
  &&  {\partial \rho_i(x,\mathbf{k},t)\over\partial t}
  =e_iE(x,t){\partial \rho_i(x,\mathbf{k},t)\over \partial k_x}
  +{k_x\over m_i^{\ast}}{
    \partial \rho_i(x,\mathbf{k},t)\over\partial x} \nonumber \\ && 
\mbox{ }\hspace{0.3cm} 
+i[\mathbf{h}_i(\mathbf{k})\cdot
{\bgreek{\sigma}\over 2},\rho_i(x,\mathbf{k},t)]
+{\partial\rho_i(x,\mathbf{k},t)\over\partial t}\Bigr|_{\mathtt{s}} \;.
\end{eqnarray}
Here we assume that the transport direction is along the
$x$-axis and the carriers only occupy the lowest subbands. 
$\rho_{i}(x,\mathbf{k},t)$ are the 
electron/hole density matrices
with momentum $\mathbf{k}=(k_x,k_y)$ at position $x$. The diagonal and
off-diagonal elements of 
$\rho_i(x,\mathbf{k},t)$ stand for electron distribution functions
$f_{i\nu}(x,\mathbf{k},t)$ and spin coherence, respectively. 
The carrier densities 
at position $x$ with spin $\nu$ $(=\pm)$ are 
$N_{i\nu}(x,t)=\sum_{{\bf k}}f_{i\nu}(x,\mathbf{k},t)$ and the 
total carrier densities are
$N_i(x,t)=\sum_{\nu}N_{i\nu}(x,t)$. 

The right hand sides of KSBEs describe the drift of electrons/holes
driven by the electric field $E(x)$, diffusion caused by the spatial
inhomogeneity, spin precession around the total magnetic field ${\bf
  h}_{i}({\bf k})$ and all the scattering, respectively. By assuming
that the effect of the background charges is compensated by the
dopants, so that there is no nonzero in-plane electric field in the
absence of excited charges, the electric field $E(x)$ is determined by
the Poisson equation 
\begin{equation}
  \label{eq:Poisson}
  {\partial_x E(x,t)}=
  \sum e_i[N_i(x,t)-N_i^0]/\epsilon\; ,
\end{equation}
with $e_i$, $\epsilon$ and $N_i^0=\sum_{\nu}N^0_{i\nu}$ 
being the carrier charge, dielectric constant and the background
carrier density, respectively.
The total magnetic field is composed of the effective magnetic field
$\mathbf{\Omega}_{e/h}(\mathbf{k})$ due to the SOC,
containing the Dresselhaus and Rashba
terms,\cite{dp,rashba:jetplett.39.78}
and the one from the Hartree-Fock term of the
carrier-carrier Coulomb interaction. 
The scattering terms consist of the contributions from all the relevant
scatterings, including the carrier-impurity,
carrier-phonon and carrier-carrier Coulomb scatterings. The 
expression for the Hatree-Fock and
scattering terms can be found in
Refs.~[\onlinecite{Wu201061}, \onlinecite{PhysRevB.69.245320}].
It is noted that the KSBEs are valid for the time shorter than the
recombination time, which is in the order of a few hundreds
pico-seconds in GaAs, since we have ignored the recombination of
the electrons and holes in the KSBEs. 

The spin and charge degree of freedom of the electrons and holes have
many different combinations. All of these combinations can be
described in our model using different parameters, including those
discussed in Ref.~\onlinecite{PhysRevLett.84.4220}.
To study the temporal evolution of SPCDW, we assume that the initial
conditions are the 
thermal nonequilibrium distribution, {\em ie.}, the Fermi
  distribution with an equilibrium temperature $T$ but
 nonequilibrium chemical potential $\mu^i_{\nu}$, with electron and hole
densities being 
\begin{equation}
  \label{eq:CarrierDensity}
  N_{i\nu}(x,0)=N^0_{i\nu}+\delta n/2
  [1+(e_i/e)\nu P]\cos (qx).   
\end{equation}
Here $\delta n$, $P$ and 
$q$ are the initial amplitude, initial spin polarization and 
wave-vector of the optically excited SPCDW, respectively.
The spin momentum along the $z$-direction for electron and hole are
then 
\begin{eqnarray}
  S_i(x,0)
  &=&[N_{i+}(x,0)-N_{i-}(x,0)]/2\nonumber \\
  &=&S^0_{i}+\delta n e_iP/(2e)\cos (qx).  
  \label{eq:SpinDensity}
\end{eqnarray}
Solving the KSBEs together with these initial conditions and periodic
boundary condition, we are 
able to obtain the temporal evolution of the SPCDW. 

\section{Analytic Results}

We first present the analytic results from the simplified version
of the KSBEs. 
By using the relaxation time approximation for the momentum and spin
relaxation, one obtains the following drift-diffusion equations for
carrier densities $N_{i\nu}$
\begin{equation}
  {\partial_t N_{i\nu}}
  =\sum_{i'\nu'}\partial_x (D^{ii'}_{\nu\nu'}\partial_x
  N_{i'\nu'}+ \sigma^{ii'}_{\nu\nu'} E)
  -{(N_{i\nu}-N_{i\bar{\nu}})\over 2\tau_s^i}\; ,
\label{DE}
\end{equation}
with $D^{ii'}_{\nu\nu'}$, $\sigma^{ii'}_{\nu\nu'}$ and $\tau^i_s$ representing the
diffusion coefficient matrix, conductivity matrix and the spin
lifetime, respectively.
The off-diagonal elements of $D$ and $\sigma$ are due to the
carrier-carrier Coulomb interaction.
For the system near equilibrium, $D$ and $\sigma$ 
have the following relation 
$D^{ii'}_{\nu\nu'}=\sigma_{\nu\nu'}^{ii'}(\partial
\mu^{i'}_{\nu'}/\partial 
N_{i'\nu'})$,\cite{PhysRevLett.110.096601,*PhysRevLett.111.136602} with 
$\mu^i_{\nu}$ being the spin-dependent chemical potentials. 
Equations (\ref{DE})  can be expressed in term of 
carrier density $N_i=N_{i+}+N_{i-}$ and spin momentum 
$S_i=(N_{i+}-N_{i-})/2$, 
\begin{eqnarray}
  {\partial_t N_i}&=&\partial_x (D^{ii'}_{cc}\partial_x N_{i'}
  +D^{ii'}_{cs}\partial_x S_{i'} + \sigma^{i}_{c}E), 
  \label{eq:Ni}\\
  {\partial_t S_i}&=&\partial_x (D^{ii'}_{sc}\partial_x N_{i'}
  +D^{ii'}_{ss}\partial_x S_{i'} + \sigma^{i}_{s}E)
  -{S_i/\tau_s^i}\, ,\label{eq:Si}
\end{eqnarray}
with $D^{ii'}_{cc}=\sum_{\nu\nu'}D^{ii'}_{\nu\nu'}/2$,
$D^{ii'}_{ss}=\sum_{\nu\nu'}\nu\nu'D^{ii'}_{\nu\nu'}/2$, 
$D^{ii'}_{cs}=\sum_{\nu\nu'}\nu'D^{ii'}_{\nu\nu'}$, 
$D^{ii'}_{sc}=\sum_{\nu\nu'}\nu D^{ii'}_{\nu\nu'}/4$, 
$\sigma^{i}_{c}=\sum_{i'\nu\nu'}\sigma^{ii'}_{\nu\nu'}$, and 
$\sigma^{i}_{s}=\sum_{i'\nu\nu'}\nu\sigma^{ii'}_{\nu\nu'}/2$. 
$D^{ii}_{cc}$, $\sigma^i_c$, $D^{ii}_{ss}$, and $\sigma^i_s$ 
are the carrier diffusion coefficient, carrier conductivity, 
spin diffusion coefficient, and spin conductivity, respectively.  
The off-diagonal elements of diffusion matrix 
$D^{ii'}_{cc}$, $D^{ii'}_{ss}$, $D^{ii'}_{sc}$ and $D^{ii'}_{cs}$ 
describe the inter-band Coulomb drag between charge currents, spin
currents and the interplay between carrier and spin diffusion,
respectively. 
These equations can be regarded as an extension of the earlier 
results\cite{PhysRevLett.110.096601,PhysRevLett.111.136602}
to include 
spin polarization and the Coulomb drag. 
By extending the results of Ref.~[\onlinecite{PhysRevB.52.14796}] to
the spin polarized electron and hole gases, one obtains the following 
trans-conductivity due to the Coulomb drag from the KSBEs
\begin{eqnarray}
  \sigma^{ii'}_{\nu\nu'}&=&
  T e_ie_i'\sum_q \int d\omega 
  {V_q^2}\sinh^{-2}[{\omega/(2T)}]
  \nonumber \\ && 
  \mbox{}\times F^i_{\nu}(q,\omega)
  F^{i'}_{\nu'}((-1)^{1-\delta_{i,i'}}q,-\omega)\; .
\end{eqnarray}
Here $F^i_{\nu}(q,\omega)
={1\over m_{i\nu}^{\ast}}
\sum_{k}\delta(\varepsilon_{i\nu}(k)-\varepsilon_{i\nu}(k+q)+\omega)
[f^0_{i\nu}(k)-f^0_{i\nu}(k+q)]
[k_x\tau_{i\nu}(k)-(k_x+q_x)\tau_{i\nu}(k+q)]\,,
$ with $f^0_{i\nu}(k)=1/[e^{(\varepsilon_{i\nu}(k)-\mu^i_{\nu})/T}+1]$
and $\tau_{i\nu}(k)$ being the Fermi function and the momentum
relaxation time, respectively.
Using the relations between diffusion coefficient and conductivity,
one can write down the off-diagonal elements of diffusion
coefficient $D^{ii'}_{sc}$ and $D^{ii'}_{cs}$, as well as
$D^{ii'}_{ss}$. 
Especially, $D^{eh}_{ss}$ which describes
the drag between the electron and hole spins reads,
\begin{eqnarray}
  && D^{eh}_{ss}=
  T \sum_q \int d\omega 
  {V_q^2 \over \sinh^{2}\omega/(2T)} 
  [F^e_{+}(q,\omega)-F^e_{-}(q,\omega)]\nonumber\\ && 
  \mbox{}\times [F^h_+(q,-\omega)(\partial\mu_+^h/\partial N_{+}^h)
  -F^h_-(q,-\omega)(\partial\mu_-^h/\partial N_{-}^h)].
  \label{eq:spin_drag}
\end{eqnarray}

These equations describe vast different transport scenarios under
different conditions. We should first focus on the case when the
Coulomb drag is not important. 
In the traditional non-polarized ambipolar transport problem, the
carrier diffusion results in the charge inbalance in space and 
induces a nonzero electric field even if there is no applied
electric field, due to the difference between 
$D_e=D^{ee}_{cc}$ and $D_h=D^{hh}_{cc}$, the electron and hole CDCs.
This induced electric field prevents the further growth of
the charge inbalance and results in the nearly neutral charge
concentration in space. 
Consequently, even though $D_e$ and $D_h$ are quite different,
the electron and hole diffusion 
are characterized by a single 
ambipolar diffusion coefficient (ADC)\cite{smith_book}
\begin{equation}
  \label{eq:Da}
  D_a=(\sigma_e D_h+\sigma_h D_e)/(\sigma_e+\sigma_h),  
\end{equation}
where
$\sigma_e=N_e^0e\mu_e$ and $\sigma_h=N^0_he\mu_h$ are the electron and
hole conductances, with $\mu_e$ and $\mu_h$ being the electron and hole
mobilities, respectively. Theoretically, this is obtained by 
removing the term involving ${\partial E/\partial x}$ 
in Eq.~(\ref{eq:Ni}) and then setting 
$N_e(x,t)-N^0_e=N_h(x,t)-N^0_h$ to enforce the charge
neutrality.\cite{smith_book}

The spin diffusion is more complex than the charge diffusion.
When both electrons and holes are fully spin polarized, such as in the
intrinsic (undoped) system where nearly all the carriers are optically
excited, one can easily see that the evolution of charges and spins
are described by the same equations. Therefore, in this special case
the spins are indeed bound to the charges. The SDCs 
for both electron and hole should also be the ADC.

However, the situation is quite different if there are unpolarized
background charges, such as in the doped nonmagnetic QWs. 
In this situation, it is not longer appropriate to follow
the stand procedure to enforce the charge neutrality for the SPCDW. 
To study the dynamics of non-fully polarized SPCDW, one has to solve
the coupled KSBEs and Poisson equations self-consistently without
artificially imposing the charge neutrality in Eq.~(\ref{eq:Ni}). 
To achieve this, we use
the periodic boundary condition and rewrite the 
simplified transport
and Poisson equations in the form of the Fourier components
for charge concentrations and spin polarization, {eg.},
$N_i(x,t)=\sum_lN^l_{i}(q,t)e^{ilqx}$. Limited 
to the linear order of excited charge concentrations, we find that
only $l=0,\pm 1$ components are relevant and obtain 
the following solution for the charge concentrations,
\begin{eqnarray}
  \label{eq:Ni1}  
  && N_{i}^{\pm 1}(q,t)=
  \delta
  n\bigl\{e^{-\lambda_-t} -(e^{-\lambda_-t}-e^{-\lambda_+t}) \nonumber\\ &&
  \mbox{}\times {\sigma_i e_i}
  {\epsilon (D_e-D_h)q^2/[e (\sigma_e+\sigma_h)^2]}
  \bigr\},
\end{eqnarray}
where to the quadratic term of $q$, 
$\lambda_{+}=(\sigma_e+\sigma_h)/\epsilon+
(\sigma_eD_e+\sigma_hD_h)q^2/(\sigma_e+\sigma_h)$
and
$\lambda_{-}=(\sigma_eD_h+\sigma_hD_e)q^2/(\sigma_e+\sigma_h)=D_aq^2$. 
The solution shows that electrons and holes indeed diffuse together as
a bundle with the ADC  
$D_a$ except a small difference that results in a small charge
imbalance,
\begin{equation}
N_{e}^{\pm 1}-N_{h}^{\pm 1}=-\delta n q^2\epsilon
[(D_e-D_h)/(\sigma_e+\sigma_h)]
(e^{-\lambda_-t}-e^{-\lambda_+t}),
\end{equation}
which induces an electric field that prevents the further growth of
the charge imbalance.

Using the above result for the charge concentrations, one then 
obtains the equations for the spin diffusion, 
\begin{eqnarray}
  && \partial_tS_{i}^{\pm 1}(q,t)=
  -\delta n {e_i\mu_i(D_e-D_h)}
  (e^{-\lambda_-t}-e^{-\lambda_+t})q^2
  \nonumber \\ && 
  \mbox{}\times S_i^0/(\sigma_e+\sigma_h)
  -D_iq^2S_{i}^{\pm 1}-S_i^{\pm 1}/\tau^i_s\, .
  \label{eq:diff-ele-spin}     
\end{eqnarray}
The first term on the RHS of
Eq.~(\ref{eq:diff-ele-spin}) is the 
generation of the SDW from spin-charge coupling between 
the spatial homogeneous spin polarization and CDW. 
The generation rates are proportional to the charge inbalance and the
homogeneous spin polarization. 
It should be noted that this SDW generation from the CDW is purely
from the orbital motion and requires initial spin polarization. This
should not be confused with the SDW generation from the CDW due to the 
SOC.\cite{PhysRevLett.111.136602}
The last two terms are diffusion and relaxation of the
SDW, respectively. 
These equations show that the SDW diffuses independently with
different SDCs $D_i$ ($i=e,h$), same as the individual CDCs. This
contradicts to all previous assumptions that the SDC should also be
the ADC 
$D_a$ [Eq.~(\ref{eq:Da})], based on the assumption that carrier spin 
degrees of freedom are bound to the charge degrees of freedom. 

The difference between the charge diffusion and spin diffusion
originates from the fact that spin diffusion does not necessarily
result in charge inbalance even when the spatial profiles of the CDW and
SDW overlap with each other, such as the case of SPCDW. For example,
for SPCDW in $n$-type QWs or SPCDW with low spin polarization, the
change in electron concentration caused by the spin-up electrons
diffusing 
from left to right can be compensated by the one caused by spin-down
electrons moving in the opposite direction. The net effect of this
process is the pure spin diffusion that reduces the spin polarization 
without any accompanied charge diffusion. 
Therefore, the spin motion can be separated from the charge
motion. In the case of SCS, the electron and hole SDCs
can be quite different from each other. 
The difference between the charge and spin diffusion can be further
enhanced by the Coulomb drag between electron and hole spin currents
when the spin polarization is high enough. 

\begin{figure}[htbp]
  \centering
  \epsfig{file=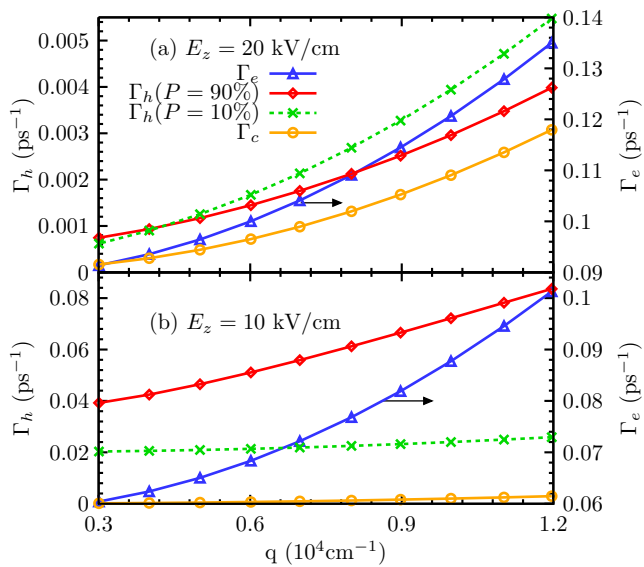,width=8.5cm}
  \caption{The decay rates $\Gamma_c$ of CDWs (Brown circle), 
    the average decay rates $\Gamma_e$ for electron SDW (Blue
    triangle), $\Gamma_h$ for hole SDW with initial polarization
    $P=90\%$ (Red diamond) 
    and $\Gamma_h$ for hole SDW with 
    $P=10\%$ (Green cross) 
    vs wave-vector
    for $n$-type GaAs $(111)$ QWs with (a) $E_z=20$~kV/cm and (b)
    $E_z=10$~kV/cm. 
    The relaxation rates for electron and hole CDWs
    are identical to each other. 
    The electron spin relaxation rates with different initial
    polarization coincide with each other. 
    Note that the scale of $\Gamma_e$ is on the right hand side of the
    frame.
  }
  \label{fig:rates}
\end{figure}

\section{Numerical results}
The above simplified KSBEs can only qualitatively describe the
temporal evolution of the SPCDW. 
Especially, Eq.~(\ref{eq:diff-ele-spin})
is valid when the spin polarization and the effect of the
Coulomb drag are small.
To justify the existence of the SCS
in genuine
situation and to study the effect of the Coulomb drag, 
we now turn to the numerical solution of
the full KSBEs, which include all the relevant scatterings, such as the
carrier-impurity, carrier-phonon as well as carrier-carrier Coulomb
scatterings, for the SPCDW in $(111)$ GaAs QWs. 
We choose the out-of-plane electric field $E_z$, which modifies 
the strength of the SOC,\cite{sun:093709,PhysRevB.85.235308} in a
regime not far away from cancellation 
point of the hole SOC, so that the spin lifetimes of electron
and hole are long enough. 
From the expressions of the SOC [Eqs.~(1,2) and (A3-A7) for holes in
Ref.~[\onlinecite{PhysRevB.85.235308}] and 
Eqs.~(1-3) for electrons in Ref.~[\onlinecite{sun:093709}]],
the cancellation point of hole SOC is $E_z=22$~kV/cm for the QW we
study. 
The numerical calculations on 
the evolution of the SPCDW are carried out
for a series of wave-vector $q$ under
different $E_z$ at 
$T=30$~K in an $n$-type GaAs QW with the background carrier densities
$N^0_e$, $N^0_h$, impurity density $N_i$ being $4\times
10^{11}$~cm$^{-2}$, 0, and $0.1~N_e^0$, respectively. 
By fitting the temporal evolutions of the CDWs and SDWs exponentially
or biexponentially, one obtains the corresponding decay rates, which
are functions of the wave-vector $q$. The decay rates can be
fitted using quadratic functions $\Gamma=Dq^2+cq+1/\tau$ to obtain the
diffusion coefficients $D$.\cite{weng:063714,PhysRevB.86.205307}

In Figs.~\ref{fig:rates}(a) and (b), we present the decay
rates of the optically excited 
SPCDWs with initial amplitude $\delta n=0.1N^0_e$ and 
spin polarizations $P=10$\% and $90$\% 
as function of the wave-vector $q$ with $E_z=20$~kV/cm 
(near the cancellation point of the hole SOC) and
$10$~kV/cm (away from the cancellation point), respectively. 
We first focus on the case of low initial spin polarization. 
We find that the numerical results for $P=10$\% quantitatively confirm
the analytical results presented in the previous section. Namely, 
the charge transports are ambipolar diffusion, 
described by the ADC
% ambipolar diffusion coefficient 
$D_a$, 
and the motions of the
charges and spins are separated. The electron SCS
%spin-charge separation
becomes obvious once one considers the difference between the electron
%spin diffusion coefficient 
SDC $D^e_s$ (about $300$~cm$^2$/s) and
$D_a$ (about $20$~cm$^2$/s). For hole SDW, even though the 
% spin diffusion coefficient 
SDC
$D^h_s$ (about $30$~cm$^2$/s) are slightly
larger than $D_a$, the 
evidence of the hole SCS
% spin-charge separation 
is marginal. However,
this is expected, since according to 
Eq.~(\ref{eq:diff-ele-spin}) $D^h_s$ should be close to the hole
CDC $D_h$, which is almost the same as $D_a$, 
with or without the SCS.

A strong but somehow surprising evidence for the hole SCS
comes from the results with high initial spin
polarization. For $P=90$\%, the hole SDC
$D^h_s$ is $24$~cm$^2$/s, which is close to the hole CDC
$D_h$, when $E_z=20$~kV/cm. As understood from the
previous section, this numerical result leads to the conclusion that
the spin and charge motions are not separatable for nearly fully
polarized holes when $E_z=20$~kV/cm. However, once 
$E_z$ drops to $10$~kV/cm, $D^h_s$ drastically increases to
$290$~cm$^2$/s, much larger than $D_h$. 
The huge difference between $D^h_s$ and $D_h$ when 
$E_z=10$~kV/cm 
is a strong indication of hole SCS. 
However, it is also
a surprising result, since one does not expect to see
that $D^h_s$ can exceed $D_h$ by one order of magnitude even in the
case of the SCS 
from Eq.~(\ref{eq:diff-ele-spin})
\footnote{
Contradictively, the initial spin polarization has marginal effect on
the dynamics of the CDWs since the spin polarization does not directly
affect the charge motion. Moreover, the existence of the large number
of the unpolarized background electrons also makes the initial spin
polarization irrelevant to the dynamics of the electron SDWs.}.

The markedly enhancement of $D^h_s$ when $E_z=10$~kV/cm 
is understood as a result of the
joint effect of SCS
% spin-charge separation 
and the Coulomb drag between
electron and hole spin currents. In our system, the hole spin motion
is bound to the carrier motion for the holes when they are excited
since the holes are almost fully polarized.  
When $E_z=20$~kV/cm, the hole spin lifetime is very long (about
1.7~ns),
therefore the 
hole spin and charge are bound together in the time
regime we study (up to a few hundred pico-seconds),
as there is not
effective compensative motion from holes with the opposite spin. 
However, for $E_z=10$~kV/cm the hole spin lifetime reduces
to 30~ps, therefore the hole spin and charge motions can be separated
for time longer than 30~ps. 
Moreover, due to the Coulomb scattering between electron and hole
spins, the fast moving electron spins can drag the hole spins to
move together with them in the case of hole SCS 
and result in the enhancement of the hole SDC.
To verify that the Coulomb drag is the reason of the markedly
enhancement of $D^h_s$, 
we perform the numerical calculations without the electron-hole
Coulomb scattering and find that indeed there is not big difference
between $D^h_s$ and $D_h$ for any $E_z$.
It should be further noted that, the spin Coulomb drag between electron
and hole is stronger for higher spin polarization.
This is the reason that $D^h_s$ is only slightly larger than $D_h$ in
the case of low spin polarization, while it can be one order of magnitude
larger than $D_h$ in the case of large spin polarization. 
One can further conclude that, the markedly enhancement of $D^h_s$ can
only be seen for 
the highly spin polarized SPCDWs in the $(111)$ GaAs QWs where both
electrons and holes have proper
spin lifetimes for the spin polarization to relax fast
enough to allow the SCS
but not too fast to
reduce the strong Coulomb drag. 

To observe this enhanced $D^h_s$ experimentally, we propose
to use the transmission-grating-photomasked transient spin grating
technique to generate a spin polarized carrier grating with
circularly polarized laser beam in the $(111)$ GaAs
QWs,\cite{oe.20.008192,*oe.20.003580}   
and monitor the evolution of the total electron and hole spin
momentum using the circular dichromatic time-resolved pump-probe
absorption spectroscopy.\cite{oe.20.008192,*oe.20.003580}  
Without the enhanced $D^h_s$ one is expected to observe fast changes in
the total spin momentum as the electron spins quickly diffuse away,
followed by a slow decay from the hole spins left behind. In the case of
the markedly enhanced $D^h_s$, one should not be able to observe any
slow decay since the hole spins diffuse almost as fast as the electron
ones. 

\section{Conclusion}
In conclusion, we study the possibility and the consequence of the
SCS when spatial inhomogeneities of charge and spin
overlap in $n$-type $(111)$ GaAs QWs microscopically. 
We analytically show that, even though the motions of electron and
hole CDWs are bound together due to the requirement of the charge
neutrality, their spin motions can be quite different from the charge
motions. Only in the intrinsic system with fully polarized SPCDW, the
spin and charge motions of both electrons and holes are bound together
and their diffusions are all characterized by the ADC.
When carriers are not fully spin polarized, the change in the
charge density caused by the spin-up carriers 
can be compensated by the opposite motion of the spin-down carriers. 
As a result, the spin and charge motions can be separated
and the SDC
can be quite different from the
ambipolar one. The full numerical calculations confirm our analytical
results that in the $n$-type QWs 
the charge motions of the electrons and holes are indeed bound
together, but the electron spin and charge diffusions are independent
on each other due to the existence of large unpolarized background
electrons.  
However, the spin and charge motions of holes can only be separated
when the initial spin polarization is small or the spin lifetime is
short. More importantly, 
when the spin and charge motions of holes can
be separated, 
the Coulomb drag between the spin polarized electrons
and holes can have dramatic effect on the spin diffusion. 
With proper initial spin polarization and electron/hole spin 
lifetimes, which can only be achieved in $(111)$ QWs, it
is possible to observe more than one-order-of-magnitude 
enhancement for hole SDC
as the joint effects of the Coulomb drag and SCS.

\begin{acknowledgments}
This work was supported by 
the National Natural Science
Foundation of China under Grant No. 11334014,
the National Basic Research Program of
China under Grant No.\ 2012CB922002 and the Strategic
Priority Research Program of the
Chinese Academy of Sciences under Grant No. XDB01000000. 
\end{acknowledgments}

%\bibliography{ref,ref_new}
%merlin.mbs apsrev4-1.bst 2010-07-25 4.21a (PWD, AO, DPC) hacked
%Control: key (0)
%Control: author (8) initials jnrlst
%Control: editor formatted (1) identically to author
%Control: production of article title (-1) disabled
%Control: page (0) single
%Control: year (1) truncated
%Control: production of eprint (0) enabled
%

\end{document}